# Formation of Ag and Au Plasmonic Nanoparticles by Ion Implantation in $Ga_2O_3$ thin films

Inês Freitas[1,2], Ana Sofia Sousa[1,2], Duarte Magalhães Esteves[1,2], Mamour Sall[3], Ângelo Rafael Granadeiro da Costa[4], Joana Madureira[4,5], Sandra Cabo Verde[4,5], Katharina Lorenz[1,2,5], Marco Peres[1,2,5]

[1] INESC Microsystems and Nanotechnology, Rua Alves Redol 9, Lisbon 1000-029, Portugal

[2] Institute for Plasmas and Nuclear Fusion, Instituto Superior Técnico, University of Lisbon, Av. Rovisco Pais 1, Lisbon 1049-001, Portugal

[3] Université Caen Normandie, CEA, ENSICAEN, CNRS, Normandie Univ, CIMAP UMR6252, F-14000 Caen, France

[4] Centre for Nuclear Sciences and Technologies, Instituto Superior Técnico, University of Lisbon, Estrada Nacional 10, km 139.7, Bobadela 2695-066, Portugal

[5] Department of Nuclear Science and Engineering, Instituto Superior Técnico, University of Lisbon, Estrada Nacional 10, km 139.7, Bobadela 2695-066, Portugal

* Corresponding authors: ana.sofia.sousa@tecnico.ulisboa.pt, duarte.esteves@tecnico.ulisboa.pt;

## Abstract

Gallium oxide ($Ga_2O_3$) is a wide-bandgap semiconductor with exceptional electrical and optical properties, making it a promising material for optoelectronic and sensing applications. In this work, we demonstrate for the first time the formation of plasmonic silver (Ag) and gold (Au) nanoparticles embedded in $Ga_2O_3$ thin films via ion implantation. $Ga_2O_3$ films deposited by RF sputtering on sapphire substrates were implanted with Ag or Au ions at 150 keV and a nominal fluence of $5 \times 10^{16}$ ions/cm$^2$, followed by thermal annealing between 200 and 700 °C. Rutherford backscattering spectrometry (RBS) measurements revealed saturation effects during implantation, resulting in lower incorporated fluences, as well as out-diffusion with post-implantation annealing. Transmission electron microscopy confirmed the formation of metallic nanoparticles with a distribution consistent with the metal profiles measured by RBS. Optical absorption measurements showed a pronounced localized surface plasmon resonance (LSPR) band in the Ag-implanted films, visible even in the as-implanted state and red-shifting with increasing annealing temperature, while Au-implanted films exhibited a distinct LSPR peak only after annealing at ≥500 °C. The observed LSPR shifts with annealing are attributed primarily to changes in the $Ga_2O_3$ matrix rather than a change in nanoparticle size. These results establish ion implantation as a viable approach for integrating plasmonic nanostructures into $Ga_2O_3$.

**Introduction**

Gallium oxide ($Ga_2O_3$) has drawn significant interest from the scientific community due to its unique optical, electrical, chemical, and thermal characteristics. This material presents a wide bandgap of approximately 4.9 eV at room temperature, tunable electrical properties ranging from nearly insulating to highly conductive, and an outstanding high breakdown electric field of 8 MV/cm [1], making it highly promising for applications in photodetection [2] and high-power electronics [3].

Plasmonic metallic nanoparticles, such as Ag and Au nanoparticles, exhibit a unique phenomenon named localised surface plasmon resonance (LSPR), in which conduction electrons collectively oscillate in response to incident light [4]. LSPR enhances local electromagnetic fields, increasing light absorption and scattering, and is highly tuneable via particle size, shape, and the refractive index of the surrounding medium [4]. These optical properties make them valuable candidates for applications in sensing [5,6], imaging [7], and energy harvesting [8]. In recent years, the integration of plasmonic nanostructures has created new opportunities for enhancing the performance of wide bandgap semiconductor-based devices, namely in the field of sensing [9]. In fact, embedding nanoparticles in semiconductors has been shown to improve and tune the photodetection responsivity [10–14], accelerate surface reactions for gas sensing [15–17], and enhance catalytic and biosensing performance [18,19].

Although the formation of Au and Ag nanoparticles by ion implantation in other semiconducting or dielectric thin films, such as $SiO_2$ [20] and $TiO_2$ [21,22], has been widely studied before, there is no record of $Ga_2O_3$ being studied for this purpose. The main advantage of ion implantation over other nanoparticle formation methods, such as the incorporation of the metal during the growth of the film [23], is the precise control over the depth profile of implanted ions and the possibility for lateral patterning. Therefore, this work aims to create Au and Ag nanoparticles by ion implantation in $Ga_2O_3$ thin films and investigate their plasmonic properties.

**Experimental details**

In this work, thin films of $Ga_2O_3$ were deposited by radio-frequency (RF) sputtering using the Alcatel SCM 450 system installed at INESC MN. They were grown at 80 W RF power, with an Ar flow rate of 20 sccm and a working pressure of approximately 3.80 mTorr, at room temperature. The amorphous deposited films were then annealed at 1000 °C for 1 h in air, using a tubular furnace, to promote crystallization [24]. These samples were grown on *c*-plane sapphire due to its optical transparency from the infrared (IR) to the ultraviolet (UV), making it ideal for optical transmission studies, as well as its thermal stability up to 1500 °C, which is essential to withstand the annealing steps. Additionally, being a crystalline substrate, *c*-sapphire has been shown to promote the crystallisation of $\beta$-$Ga_2O_3$ during thermal annealing [25].

The ion implantation was performed using a Danfysik Model 1090 high-current ion implanter, installed at the Laboratory of Accelerators of Instituto Superior Técnico (IST) [26]. The $Ga_2O_3$ films were implanted with either Ag or Au at 150 keV, with a nominal fluence of $5 \times 10^{16}$ ions/cm$^2$. Then, the implanted samples were annealed at temperatures between 200 and 700 °C for a systematic study of the effects of the annealing temperature on the plasmonic properties of the nanoparticles.

Rutherford backscattering spectrometry (RBS) measurements were carried out using the 2.5 MV Van de Graaff accelerator installed in IST [26] with a 2 MeV He$^+$ beam, and a beam current of approximately 2 nA. A total charge of 2 µC was acquired per spectrum. The signal was collected simultaneously using three Si PIN diodes detectors placed at backscattering angles of 140° and ±165°. The experimental data from all three was then simultaneously analysed using the NDF code [27], which enables an evaluation of the thickness and stoichiometry of the films, as well as the depth-profile of the implanted ions.

Optical properties were investigated using a Shimadzu UV-1800 spectrophotometer at the Laboratory of Technological Assays in Clean Rooms of Instituto Superior Técnico. This setup includes a halogen lamp for longer wavelengths, a deuterium lamp for shorter wavelengths, and an integrated monochromator. The measured transmitted light intensity is automatically normalised to that of a reference incident beam, allowing for the absorbance to be directly obtained for wavelengths from 190 to 1100 nm.

The structural properties of the samples were also analysed through scanning transmission electron microscopy (STEM), performed at the Center of Research on Ions Materials and Photonics (CIMAP) in Caen. Thin foils were cut out from the implanted samples in a cross-sectional configuration, using a 30 keV focused Ga$^+$ ion beam (FIB) in a Helios Nanolab 660 system (Thermofisher). The beam current was gradually reduced in several steps, from 9 nA to 80 pA, to achieve samples with a thickness of approximately 100 nm. A final cleaning step was performed by gradually decreasing the FIB voltage to 2 kV and 1 kV with a current of 23 pA and 28 pA, respectively. STEM observations were performed on a double-corrected JEOL ARM 200F equipment operating at 200 kV, equipped with a GIF Quantum Gatan energy filter. For imaging, we used a high-angle annular dark field (HAADF) detector at inner and outer angles of 68 and 280 mrad, respectively.

**Results and Discussion**

RBS was performed on the samples before and after ion implantation, as well as after post-implantation annealing, to study their composition and depth profile. The resulting spectra are represented in Figure 1a along with the respective fits. The total amounts of Ag and Au obtained from this RBS analysis were $3.9 \times 10^{16}$ at/cm$^2$ and $1.7 \times 10^{16}$ at/cm$^2$, translating to a total amount of 9.8 at.% for Ag and 4.8 at.% for Au for the implanted layer. The actual implantation fluences were lower than the nominal values, which could suggest saturation of the host matrix, meaning that increasing the fluence further causes sputtering instead of increasing the total implanted amount [22,28]. Sahu *et al.* [29] reported this effect, implanting Au and Ag ions on host matrices of silicon and silica glass, at fluences above $3 \times 10^{16}$ at/cm$^2$. They also observed a higher retention of Ag vs Au with increasing fluence, which is in agreement with our results. However, we then observed a decrease in these incorporated ions to $1.8 \times 10^{16}$ at/cm$^2$ and $1.7 \times 10^{16}$ at/cm$^2$ for Ag and Au, respectively, after annealing at 500 °C. This type of behaviour has been previously reported and assigned to out-diffusion [30,31].

The implanted ion profile can be extracted from this analysis, and is represented in Figures 1b and 1c. The maxima of the Ag and Au concentration profiles occur at depths of ~22 and ~17 nm, respectively. To evaluate these results, Stopping Range of Ions in Matter (SRIM) [32] Monte Carlo simulations were performed, considering a density of 5.88 g/cm$^3$ [1] and displacement threshold energies for Ga and O atoms of 28 and 14 eV, respectively [33]. As shown in Figures 1b and 1c, these showed expected ranges of 38 and 30 nm, respectively, as well as displacements profiles with maxima at 25 and 18 nm, respectively. It appears that the real profile more closely follows that of the damage profile rather than the expected range distribution. This is in agreement with the results reported by Jagerová *et al.* [34] in Au-implanted GaN, who showed that these high fluences can lead to diffusion towards regions with higher disorder, closer to the surface. On the other hand, simulations performed by Stepanov *et al.* [35], show a similar effect with increasing fluence, due to sputtering. The implanted samples feature a thinner $Ga_2O_3$ film than the as-grown sample (Fig. 1a), which may suggest sputtering. We cannot, however, disregard lateral inhomogeneities stemming from the RF Sputtering.

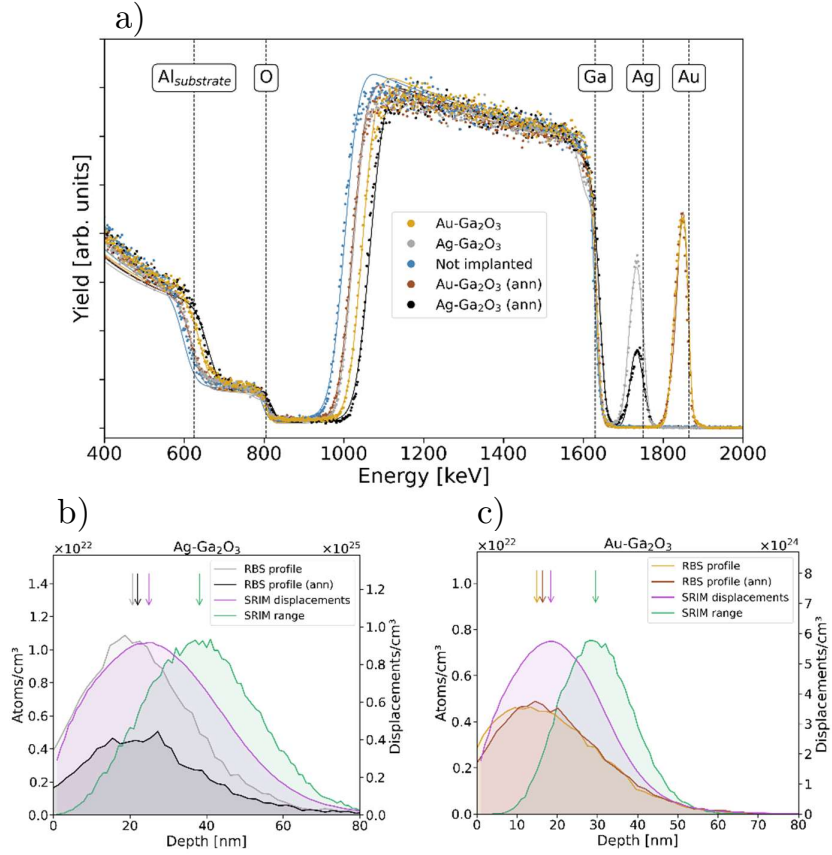

**Fig. 1** - RBS spectra and respective fits of $Ga_2O_3$ thin films, from the detector at 140°, before and after being implanted, and then after being annealed (ann), where the surface energies of Ga, Ag and Au are identified (a). Concentration profiles of the implanted ions for Ag- (b) and Au-implanted $Ga_2O_3$ (c) obtained from NDF fits of the RBS data, compared to the SRIM simulation, using the fluence estimated from the RBS fits.

The XRD $2\vartheta/\omega$ scans presented in Figure 2 suggest that, in agreement with our previous work [25], this sample is polycrystalline and textured with a preferential ($\bar{2}01$) orientation, as the relative peak intensities differ from those of the powder pattern [36]. Considering the volume sensitivity of the technique, no peaks related with other phases have been observed. However, it has been reported in the literature that this material undergoes an amorphous to $\gamma$-phase transition during the initial annealing, before reaching the $\beta$ phase [37]. A $\beta$- to $\gamma$-phase transition also is known to occur due to ion implantation in bulk samples, already at damage levels below those incurred here (0.78 dpa vs 97 and 61 for Ag and Au, respectively) [38], but it was not detected in our samples. After the post-implantation annealing at 500 °C, only the 111 reflection of Ag is tentatively identified, suggesting that the introduced nanoparticles may be too small to give rise to significant diffraction peaks. Moreover, some of the expected peaks (e.g. Ag 220) could be overlapped with peaks from the host matrix (e.g. $\bar{7}12$).

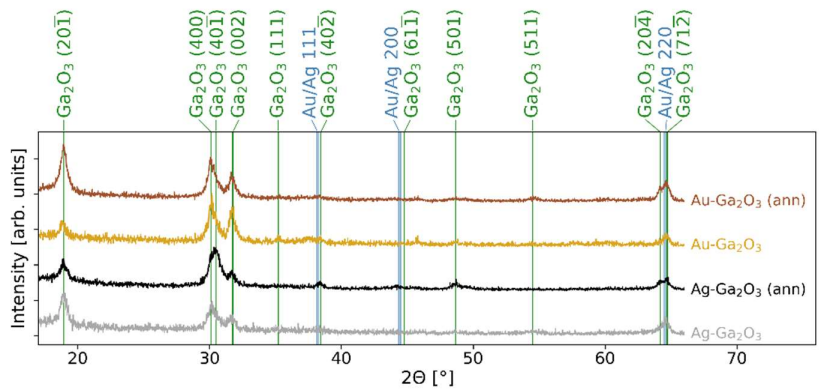

**Fig. 2** - 2ϑ/ω scans of $Ga_2O_3$ thin films, as-implanted and after thermal treatment at 500 °C (ann). The most likely peaks pertaining to the matrix are marked in green, while the probable ranges for nanoparticle peaks are marked in blue.

In order to gain more insight on the formation, distribution and morphology of the nanoparticles induced by ion implantation, as-implanted and annealed samples were analysed using transmission electron microscopy. This technique allowed us to observe the morphology and spatial distribution of the implanted nanoparticles within the matrix, as seen in Figure 3. The ion concentration profiles from the RBS fits are shown overlaying these images, for easier comparison.

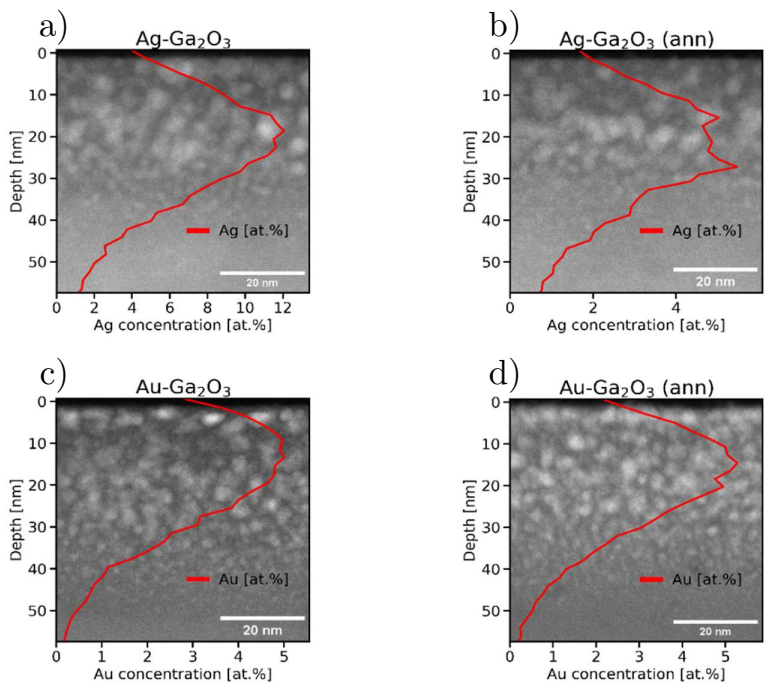

**Fig. 3** – STEM-HAADF images showing the profiles of Ag-implanted $Ga_2O_3$ thin films, both as-implanted (a) and annealed at 500 °C (b), and Au-implanted $Ga_2O_3$ thin films, both as-implanted (c) and annealed at 500 °C (d).

TEM results revealed the presence of metallic nanoparticles from the surface up to a depth of 42 nm and 50 nm in the Ag- and Au-implanted samples, respectively, and no significant changes are seen in shape, size or distribution after annealing at 500 °C. For both ions, at least two different populations of nanoparticles were identified. In the Ag sample, the bigger nanoparticles have a diameter around 5 nm and only go to a depth of about 25 nm from the surface, while the smaller nanoparticles have a diameter around 1.5 nm. In the Au sample, the bigger nanoparticles have a diameter around 3 nm and only go to a depth of about 30 nm from the surface, while the smaller nanoparticles have a diameter around 1 nm. This is in agreement with the results reported by Stepanov *et al.* [39], who also observed the formation of two distinct layers of Ag nanoparticles in silica glass, at different depths and with different sizes, considering implantation temperatures of 50–60 °C, albeit not at 20 °C. As the implantation sample plate is not actively cooled, it possible likely that temperatures above room temperature were reached.

The depth concentration profiles from the RBS data fittings generally align with the observations from the TEM analysis, confirming that ion implantation is a reliable method for controlled nanoparticle formation. In the case of Ag, the RBS ion profile has a maximum at 21 nm, and the TEM image (Figure 3a) shows that the nanoparticles of maximum size concentrate at approximately 20 nm depth, where Ag concentration is highest. Some larger nanoparticles can also be seen at the surface where Ag concentration should be low, suggesting possible surface segregation or enhanced nucleation at the surface. The same is true for the Au as-implanted and annealed samples, for which the maximum of the profile at 15 nm is roughly at the centre of the layer with the largest nanoparticles. These images also show a higher concentration of nanoparticles near the surface, which is coherent with the already mentioned saturation and sputtering.

The high-resolution STEM (HR-STEM) Au-implanted samples obtained also suggest that the formed nanoparticles show a good crystalline quality, with a well-ordered atomic arrangement. This is shown in Fig. 4, where the fast Fourier transform (FFT) of one of the nanoparticles is shown to be consistent with the expected diffraction pattern for the [110] zone axis for face-centred cubic (fcc) Au nanoparticles (Fig. 4b-d), with a lattice constant of ∼4.1 Å [40].

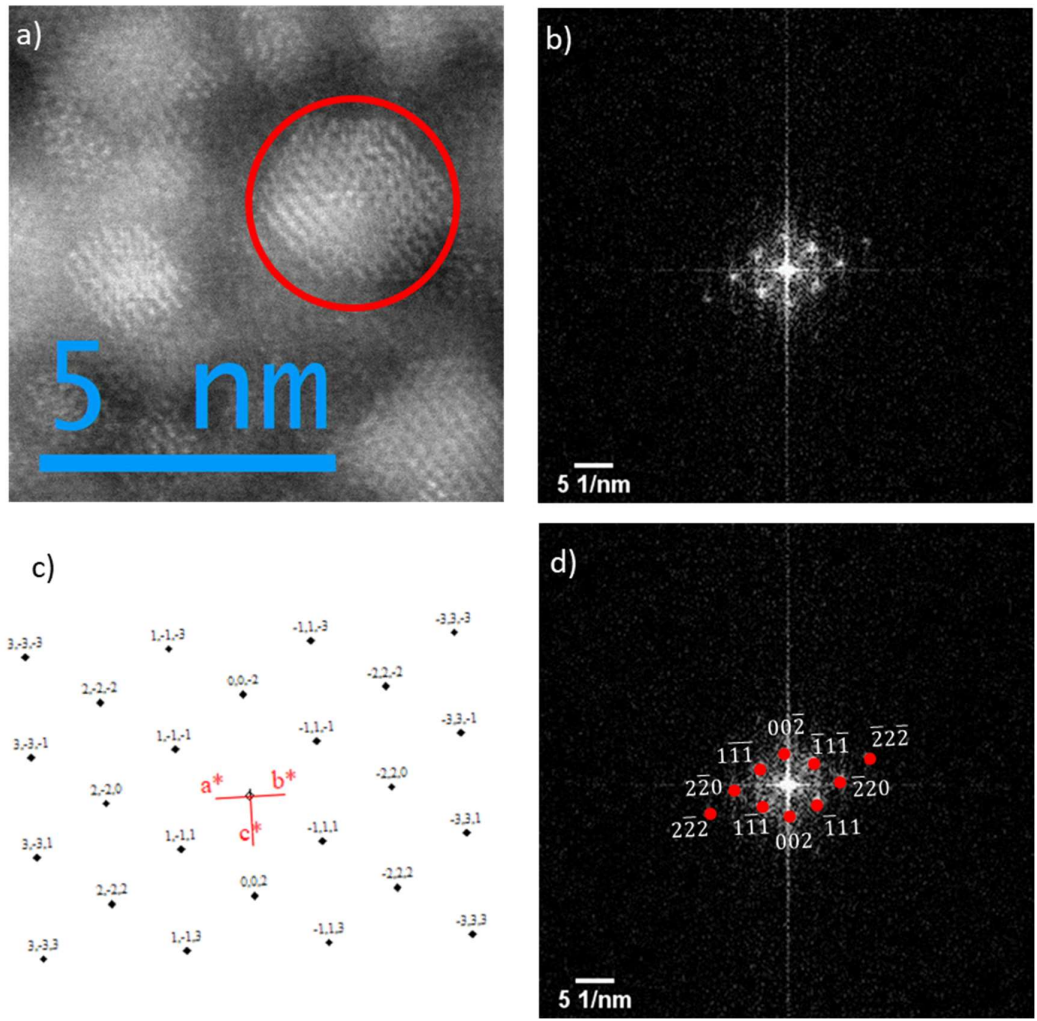

**Fig. 4** – (a) HRTEM image of the annealed sample showing crystalline Au nanoparticles, as well as (b) the FFT corresponding to the nanoparticle marked with a red circle. Panel (c) shows the simulated diffraction pattern expected for the [110] zone axis for fcc Au, while panel (d) shows panel (b) superimposed with the identified reflections.

Optical absorption measurements were carried out in order to study the optical response of the implanted samples and assess their potential plasmonic behaviour. The resulting spectra show noticeable differences depending on the nature of the implanted ion, as can be seen in Figure 5.

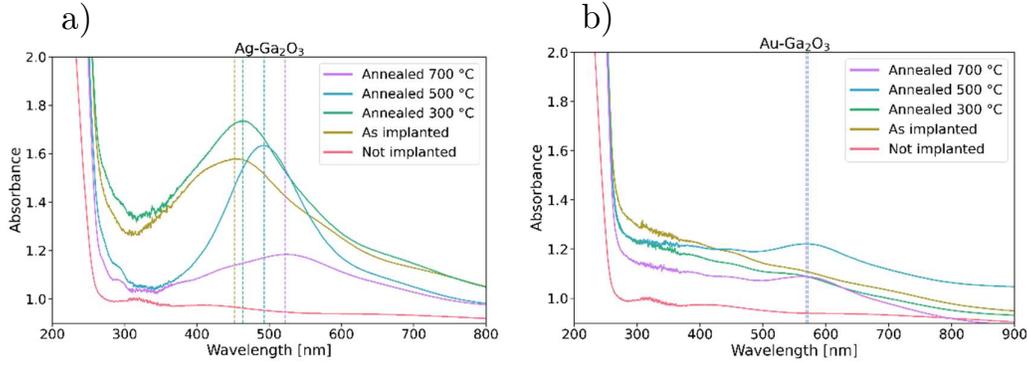

**Fig. 5** - Optical absorption spectra of $Ga_2O_3$ implanted with Ag (a) and Au (b), as-implanted, and annealed at 300, 500 and 700 °C after implantation. The spectrum of a non-implanted $Ga_2O_3$ film was also included for reference. The noise seen at 300–350 nm is an artefact created by the lamp transition.

The behaviour of the nanoparticles is highly dependent on their shape, size and distribution, but also on the properties of the surrounding matrix [41]. Therefore, the damage caused by the initial implantation has a non-negligible impact, and is mitigated as the annealing temperature increases and crystallinity recovers [42]. These effects can be clearly seen in Figures 5a and 5b, as the bandgap of the material decreases after implantation, and pronounced Urbach tails appear due to the introduced disorder, and are somewhat attenuated by annealing.

Therefore, and as shown in Fig. 5a, it is of special interest that the Ag nanoparticles show a response even in the as-implanted sample, which is improved with annealing up to 300 °C, and then deteriorates with further annealing. This is coherent with the reported [30] and observed diffusion of Ag after annealing at 500 °C. A band at this wavelength has been attributed to the LSPR of Ag nanoparticles formed after implantation into $SiO_2$ [43,44]. The Au response is much less intense, which is consistent with the measured lower incorporated fluence and smaller nanoparticle size, as determined from TEM. According to Mie theory, the prevalent process for nanoparticles of this size is absorption, which is expected to scale with $R^3$, where $R$ is the radius of a spherical nanoparticle [41]. Furthermore, it is known that Ag nanoparticles yield sharper and more intense plasmon resonance peaks, which are more susceptible to refractive index variations than those from Au NPs; this happens as Ag's real dielectric function changes more significantly over the visible spectrum and has a lower imaginary part, especially in the 400–500 nm range [41].

By analysing Fig. 3, a careful comparison between the pictures from before and after annealing suggests that there was no visible evolution of the distribution of the nanoparticles induced by ion implantation. Therefore, the shift seen in the LSPR peak with increasing annealing temperature (Figure 3a) is most likely due to changes in the $Ga_2O_3$ matrix. The plasmonic peak frequency is expected to increase with the density of the material [45], as

$$\omega_p{}^2 = \frac{Ne^2}{\varepsilon_0 m_e} \Longrightarrow \omega_p \propto \sqrt{N},$$

where $\omega_p$ is the plasma angular frequency, $N$ is the number density of electrons, $\varepsilon_0$ is the permittivity of free space and $m_e$ is the electron effective mass. Since the frequency of the peak is decreasing with the annealing temperature, this suggests a decrease in the material's density. Typically, the density of a crystalline material is expected to decrease during ion implantation due to the introduction of defects [46], while for amorphous materials it is expected to increase due to compaction [47]. As previously mentioned, ion implantation is expected to cause a transition to the $\gamma$-phase, which is slightly denser than the $\beta$-phase and has a higher (average) effective electron mass [48], meaning that a reversal of this phase transition would contribute to a decrease of the plasma frequency and thus an increase in the wavelength of the plasmonic absorption peak. Although this phase could not be detected by XRD measurements, the presence of small localized clusters cannot be discarded.

In general, structural enhancement or densification of a thin film results in an increase of its dielectric constant [49,50], which impacts the behaviour of the embedded nanoparticles [41]. The dielectric function of Au and Ag is also susceptible to change, and has been shown to affect the plasmonic absorption peak [51]. In $TiO_2$ thin films, annealing has been shown to cause a red-shift, as well as increase the overall absorption of the material due to an increase in the refractive index of the material [52]. This is in agreement with our observed red-shift, as it has been shown that the refractive index of $Ga_2O_3$ thin films increases with crystalline quality [53].

These results are quite interesting, as they show that thermal annealing is not strictly necessary in order to produce Ag nanoparticles with measurable LSPR response. On the other hand, this thermal processing is essential for the improvement of the optical characteristics of plasmonic Au nanoparticles within the host matrix.

**Conclusions**

In this study, Ag and Au nanoparticles were successfully synthesised via ion implantation within $Ga_2O_3$ RF-sputtered thin films, as confirmed by optical absorption and TEM analyses.

The RBS analysis revealed that the maxima of the concentration profiles of the incorporated Ag and Au were slightly shifted toward the surface compared to the SRIM simulations. Additionally, the actually-incorporated amounts of Ag and Au were lower than the nominal values. These saturation effects can be ascribed to a combination of diffusion and sputtering.

TEM images showed two populations of nanoparticles with different sizes and depth distributions in both Ag- and Au-implanted samples. The Ag and Au nanoparticles size and concentration aligned well with the metal depth profiles from RBS analysis, save for an unexpected surface accumulation, suggesting possible migration or enhanced nucleation at the surface, which is not resolved in the RBS spectra due to the limited depth resolution. Moreover, the crystallinity of the nanoparticles was also confirmed by FFT of the TEM data.

An absorption band was observed in the Ag as-implanted film spectrum around 450 nm, which can be attributed to the LSPR of Ag nanoparticles. This band was further improved by annealing at 300 °C, but then deteriorated after annealing at higher temperatures, presumably due to the Ag diffusion/evaporation suggested by RBS. In contrast, the Au as-implanted films did not exhibit a distinct plasmonic response, likely due to the reduced concentration of the element. An LSPR peak appeared after annealing at 300 °C, showing that thermal treatment is critical for the full integration of the Au nanoparticles in the matrix. Optical absorbance also revealed that ion implantation with both Ag and Au induced a reduction in the optical bandgap of $Ga_2O_3$, explained by defect formation and lattice disorder introduced by the implanted ions.

Overall, the results demonstrate for the first time that ion implantation is a viable technique for embedding plasmonic nanoparticles in $Ga_2O_3$ thin films.

**Acknowledgements**

The authors acknowledge funding of the Research Unit INESC MN from the Fundação para a Ciência e a Tecnologia (FCT) through Plurianual financing (UIDB/05367/2025, UID/PRR/5367/2025 and UID/PRR2/05367/2025) [doi: https://doi.org/10.54499/UIS/PRR/05367/2025 and https://doi.org/10.54499/UID/PRR2/05367/2025], as well as via the IonProGO project (2022.05329.PTDC, http://doi.org/10.54499/2022.05329.PTDC) and via the INESC MN Research Unit funding (UID/05367/2020) through Pluriannual BASE and PROGRAMATICO financing. D. M. Esteves (2022.09585.BD, https://doi.org/10.54499/2022.09585.BD) and A. S Sousa (2025.04778.BD) thank FCT for their PhD grants. The TEM and FIB experiments received funding from the ANR "Investissements d'avenir" ANR-11-EQPX-0020 (Equipex GENESIS) and the CNRS Federation IRMA - FR 3095.